\newif\ifpreprint

\preprintfalse 

\ifpreprint
\documentclass[journal=jctcce,manuscript=letter]{achemso}
\else
\documentclass[journal=jctcce,manuscript=letter,layout=twocolumn]{achemso}
\fi

\usepackage[T1]{fontenc} 

\usepackage{amsmath}
\usepackage{newtxtext,newtxmath}

\usepackage{graphicx}
\usepackage{dcolumn}
\usepackage{braket}
\usepackage{multirow}
\usepackage{threeparttable}
\usepackage{xspace}
\usepackage{verbatim}
\usepackage[version=4]{mhchem} 
\usepackage{comment}
\usepackage{color,soul}

\usepackage{mathtools}
\usepackage[dvipsnames]{xcolor}
\usepackage{xspace}
\usepackage{ifthen}

\usepackage{qcircuit}

\usepackage{graphicx,longtable,dcolumn,mhchem}
\usepackage{rotating,color}
\usepackage{lscape}
\usepackage{amsmath}
\usepackage{dsfont}
\usepackage{soul}
\usepackage{physics}
\newcolumntype{d}{D{.}{.}{-1}}

\newcommand{\ra}{\rightarrow}
\newcommand{\pis}{\pi^\star}
\newcommand{\sigmas}{\sigma^\star}
\newcommand{\pp}{\pi\ra\pis}
\newcommand{\ps}{\pi\ra\sigmas}
\newcommand{\np}{n\ra\pis}
\newcommand{\nsp}{n/\sigma\ra\pis}

\newcommand{\Td}{\%T_1}

\newcommand{\Pop}{6-31+G(d)}
\newcommand{\AVDZ}{\emph{aug}-cc-pVDZ}
\newcommand{\AVTZ}{\emph{aug}-cc-pVTZ}

\newcommand{\TZ}{cc-pVTZ}

\usepackage[colorlinks = true,
            linkcolor = blue,
            urlcolor  = black,
            citecolor = blue,
            anchorcolor = black]{hyperref}

\definecolor{goodorange}{RGB}{225,125,0}
\definecolor{goodgreen}{RGB}{5,130,5}
\definecolor{goodred}{RGB}{220,50,25}
\definecolor{goodblue}{RGB}{30,144,255}

\newcommand{\note}[2]{
\ifthenelse{\equal{#1}{F}}{
\colorbox{goodorange}{\textcolor{white}{\footnotesize \fontfamily{phv}\selectfont #1}}
    \textcolor{goodorange}{{\footnotesize \fontfamily{phv}\selectfont #2}}\xspace
}{}
\ifthenelse{\equal{#1}{R}}{
\colorbox{goodred}{\textcolor{white}{\footnotesize \fontfamily{phv}\selectfont #1}}
    \textcolor{goodred}{{\footnotesize \fontfamily{phv}\selectfont #2}}\xspace
}{}
\ifthenelse{\equal{#1}{N}}{
\colorbox{goodgreen}{\textcolor{white}{\footnotesize \fontfamily{phv}\selectfont #1}}
    \textcolor{goodgreen}{{\footnotesize \fontfamily{phv}\selectfont #2}}\xspace
}{}
\ifthenelse{\equal{#1}{M}}{
\colorbox{goodblue}{\textcolor{white}{\footnotesize \fontfamily{phv}\selectfont #1}}
    \textcolor{goodblue}{{\footnotesize \fontfamily{phv}\selectfont #2}}\xspace
}{}
}

\usepackage{titlesec}

\usepackage[fontsize=11pt]{scrextend}
\titleformat{\section}
{\normalfont\sffamily\bfseries\color{Blue}}
{\thesection.}{0.25em}{\uppercase}

\titleformat{\subsection}
{\normalfont\sffamily\bfseries}
{\thesubsection}{0.25em}{}

\titleformat{\subsubsection}
{\normalfont\sffamily}
{\thesubsubsection}{0.25em}{}

\titleformat{\suppinfo}
{\normalfont\sffamily\bfseries}
{\thesubsection}{0.25em}{}

\titlespacing*{\section}{0pt}{0.5\baselineskip}{0.01\baselineskip}
\titlespacing*{\subsection}{0pt}{0.125\baselineskip}{0.01\baselineskip}
\titlespacing*{\subsubsection}{0pt}{0.125\baselineskip}{0.01\baselineskip}

\author{Pierre-Fran\c{c}ois Loos}
	\affiliation[LCPQ, Toulouse]{Laboratoire de Chimie et Physique Quantiques, Universit\'e de Toulouse, CNRS, UPS, France}
	\email{loos@irsamc.ups-tlse.fr}
\author{Denis Jacquemin}
	\email{Denis.Jacquemin@univ-nantes.fr}
	\affiliation[CEISAM, Nantes]{Universit\'e de Nantes, CNRS,  CEISAM UMR 6230, F-44000 Nantes, France}
		
\let\oldmaketitle\maketitle
\let\maketitle\relax
     \title{A Mountaineering Strategy to Excited States: Highly-Accurate Energies and Benchmarks for Bicyclic Systems}
\date{\today}

\begin{tocentry}
	\centering
	\includegraphics[width=.45\textwidth]{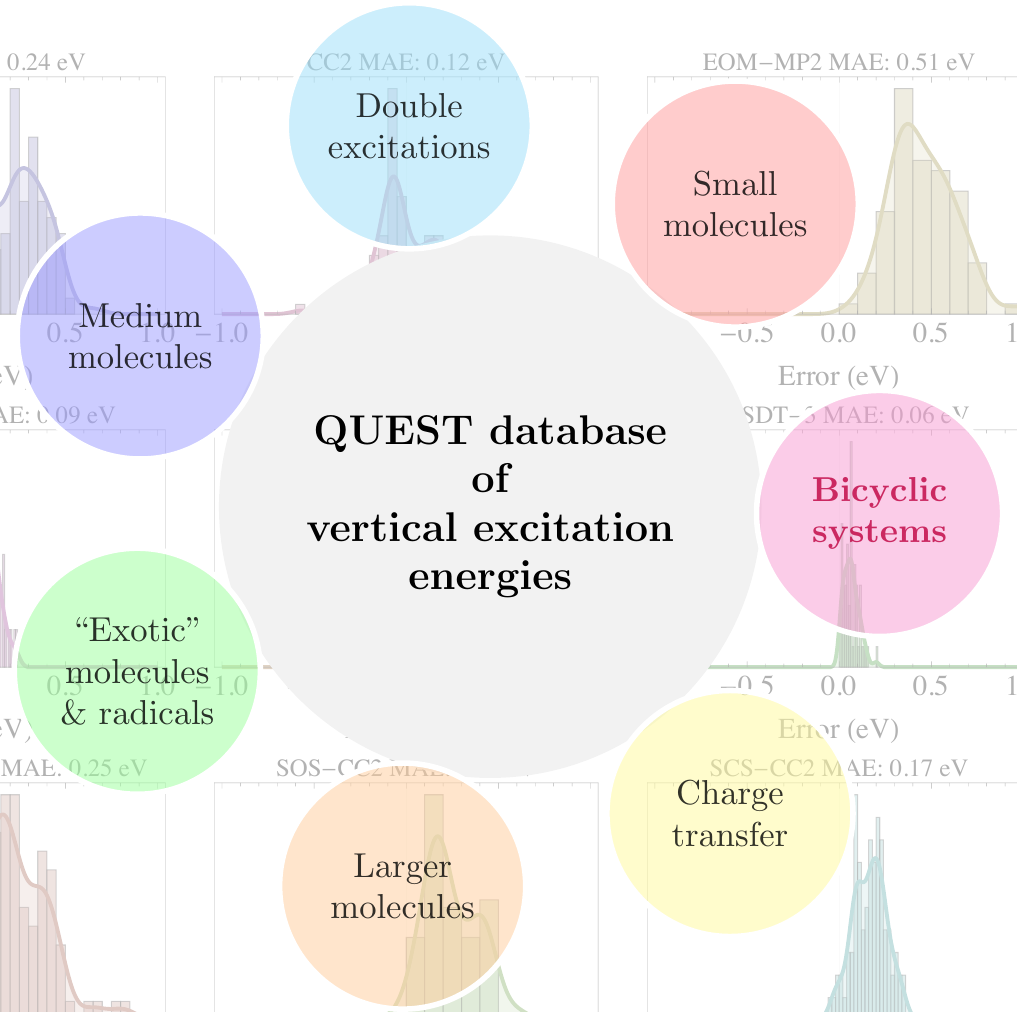}
\end{tocentry}

\begin{document}

\ifpreprint
\else
\twocolumn[
\begin{@twocolumnfalse}
\fi
\oldmaketitle

\begin{abstract}
Pursuing our efforts to define highly-accurate estimates of the relative energies of excited states in organic molecules, we investigate, with coupled-cluster methods including iterative triples (CC3 and CCSDT), the 
vertical excitation energies of 10 bicyclic molecules (azulene,  benzoxadiazole, benzothiadiazole, diketopyrrolopyrrole, fuofuran, phthalazine, pyrrolopyrrole, quinoxaline, tetrathiafulvalene, and thienothiophene). In total,
we provide \emph{aug}-cc-pVTZ reference vertical excitation energies for 91 excited states of these relatively large systems.  We use these reference values to benchmark various wave function methods, i.e., CIS(D), 
EOM-MP2, CC2, CCSD, STEOM-CCSD, CCSD(T)(a)*, CCSDR(3), CCSDT-3, ADC(2), ADC(2.5), ADC(3), as well as some spin-scaled variants of both CC2 and ADC(2). These results are compared to those obtained 
previously on smaller molecules. It turns out that while the accuracy of some methods is almost unaffected by system size, e.g., CIS(D) and CC3, the performance of others can significantly deteriorate as the systems grow, 
e.g., EOM-MP2 and CCSD, whereas others, e.g., ADC(2) and CC2, become more accurate for larger derivatives.
\end{abstract}

\ifpreprint
\else
\end{@twocolumnfalse}
]
\fi

\ifpreprint
\else
\small
\fi

\noindent

\begin{figure*}[htp]
\centering
 \includegraphics[width=.8\linewidth]{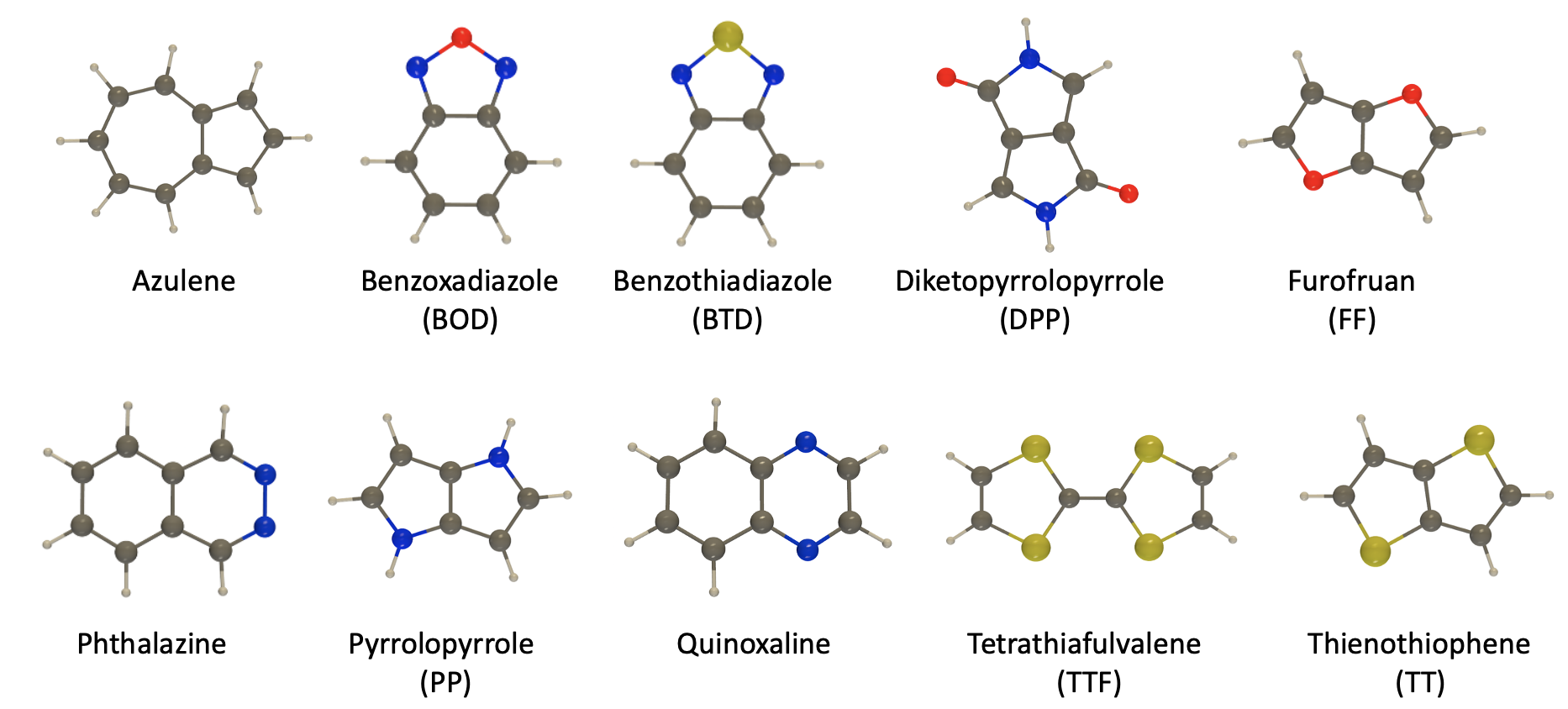}
  \caption{Representation of the investigated derivatives.}
   \label{Fig-1}
\end{figure*}

\section{Introduction}

Reference sets encompassing accurate experimental and/or theoretical data have always enjoyed a high popularity in the electronic structure community. Indeed, these allow for rapid and fair comparisons between theoretical models for 
properties of interest. The pioneering reference sets are likely the so-called Gaussian-$x$ databases originally collected by Pople and collaborators 25 years ago and constantly extended since then. \cite{Pop89,Cur91,Cur97b,Cur98,Cur07b} 
These datasets contain a large number of experimental reference values (atomization energies, bond dissociation energies, etc) and remain popular today to assess the performances of new exchange-correlation functionals (see, for example, Ref.~\citenum{Sun15a}) within density-functional theory (DFT) as well as higher-level methods. \cite{Pet12,Sce20,Yao20} Many other sets collecting reliable experimental and/or theoretical values have been developed throughout the years. The
panel is so wide that being exhaustive is beyond reach, but we can cite: (i) the S22 and S66 benchmark sets of interaction energies for weakly-bound systems; \cite{Jur06,Rez11} (ii) the HEAT set collecting high-accuracy theoretical 
formation enthalpies; \cite{Taj04} (iii) the $GW$100 and $GW$5000 sets of ionization energies; \cite{Van15,Stu20} and (iv) the very extended GMTKN$xy$ thermochemical and kinetic databases proposed by Goerigk and Grimme. 
\cite{Goe10,Goe11,Goe17} In this framework, one can certainly also pinpoint the successes of Barone's group in deriving very accurate ``semi-experimental'' structural, rotational, and vibrational reference parameters for a large panel of 
molecules of various sizes. \cite{Pic15,Pen16,Men17} 

Since 2018, our groups have made joint efforts to produce highly-accurate energies of electronically excited states (ESs) in molecular systems, \cite{Loo18a,Loo19c,Loo20a,Loo20d,Loo21a} in line
with the earlier contributions from the Mulheim group. \cite{Sch08,Sil10b,Sil10c} Our key results were recently collected in the so-called QUEST database, \cite{Ver21} that contains more than 500 theoretical 
best estimates (TBEs) of vertical transition energies (VTEs) computed in small- and medium-sized organic molecules with Dunning's {\AVTZ} basis set.  For the smallest systems (1--3 non-hydrogen atoms), 
\cite{Loo18a,Loo20d} most VTEs are of full configuration interaction (FCI) quality and they have been computed with the \textit{Configuration Interaction using a Perturbative Selection made Iteratively} (CIPSI) method. 
\cite{Hur73,Gin13,Gin15,Gar17b,Gar18,Gar19} For the molecules containing 4 non-hydrogen nuclei, \cite{Loo20a} most reference values are derived from basis-set-extrapolated coupled-cluster (CC) calculations including contributions from the singles, 
doubles, triples, and quadruples (CCSDTQ) \cite{Kuc91}. Finally for the larger systems, \cite{Loo20d,Ver21} CC with singles, 
doubles, and triples (CCSDT) \cite{Nog87,Scu88,Kuc01,Kow01,Kow01b} values are employed to define the TBEs, typically 
by computing the CCSDT excitation energies with a double-$\zeta$ basis set and correcting for basis set effects thanks to VTEs determined with the corresponding approximate third-order CC (CC3) model. \cite{Chr95b,Koc95}  

The original QUEST database contains a reasonably broad panel of organic and inorganic molecules (closed- and open-shell compounds, cyclic and linear systems, pure hydrocarbons, heteroatomic structure, etc) and ESs 
(valence and Rydberg transitions, singlet, doublet, and triplet excitations, with or without a significant double excitation character, etc) but it clearly lacked charge-transfer (CT) excitations, an aspect that we have recently corrected. \cite{Loo21a}
As the VTEs cannot be measured experimentally, but are the ``simplest'' ES property to compute, the QUEST database is especially useful for performing cross-comparisons between 
computational models. \cite{Loo20c} As examples, the VTEs included in QUEST and their corresponding TBEs have allowed us and others to (i) clearly determine the relative accuracies of CC3 and CCSDT-3; \cite{Loo18a,Loo20a} 
(ii) unambiguously define the performance of ADC(3); \cite{Loo20b} (iii) assess the fourth-order approximate CC approach (CC4) compared to its CCSDTQ parent; \cite{Loo21b} and (iv) benchmark hybrid and double-hybrid 
exchange-correlation functionals \cite{Cas19,Cas21b,Mes21,Mes21b,Gro21} as well as orbital-optimized excited-state DFT calculations. \cite{Hai21} 

Understandably, most molecules included in QUEST are rather compact. To date, only four molecules from the QUEST dataset contain 8 non-hydrogen atoms or more: octatetraene, the highly-symmetric ($D_{2h}$) benzoquinone, 
naphthalene, and tetra-azanaphthalene. \cite{Ver21} For comparison, QUEST includes 20 (10) molecules containing 4 (6) non-hydrogen nuclei. \cite{Ver21} There is a clear imbalance here, and the present contribution is a step towards  
correcting this bias by providing TBEs for a large number of ESs for the ten compounds sketched in Figure \ref{Fig-1} which encompass from 8 to 10 non-hydrogen atoms. These ten compounds have been selected to be of chemical interest 
with many building blocks used as chromophores in real-life dye chemistry. As we detail below, beyond extending the QUEST database, the present work also provides, for the vast majority of ESs considered, what can likely be 
viewed as the most accurate VTEs to date for these systems. Consequently, the data collected here can also be of interest for more specific studies, e.g., for choosing an appropriate exchange-correlation functional for studies focussing on 
optimizing the absorption properties of diketopyrrolopyrroles. 

\section{Computational details}

We closely follow the computational protocols used in our previous works, \cite{Loo18a,Loo19c,Loo20a,Loo20b,Loo20d,Ver21,Loo21a} which is briefly summarized below. 
Note that all calculations are performed within the frozen-core approximation.

The ground-state geometries are optimized at the CC3 level \cite{Chr95b,Koc95} with the {\TZ} basis set using CFOUR2.1. \cite{Mat20} These optimizations are achieved in a Z-matrix format, so that point group symmetry 
($C_{2v}$, $C_{2h}$, or $D_{2h}$) is strictly enforced. Default convergence parameters are applied. Cartesian coordinates are provided in the Supporting Information (SI) for all molecules treated here. These 
geometries are next used to compute VTEs with CC methods. Note that, in the following, we do not specify the EOM (equation-of-motion) or LR (linear-response) prefix in the CC schemes as the two formalisms 
deliver the same transition energies (yet different properties).

In a first step, we perform CCSD/{\AVTZ} calculations \cite{Pur82,Scu87,Koc90b,Sta93,Sta93b} for the 6--20 lowest-energy ESs of all molecules considering both the singlet and triplet manifolds. The main purpose 
of these calculations, performed with GAUSSIAN 16, \cite{Gaussian16} is to screen the various ESs and identify the key molecular orbitals (MOs) associated with these transitions (extended data are given in the SI 
for all molecules).  The specific nature of these ESs is determined by examining these MOs, which allows us, in the vast majority of the cases, a straightforward identification of the Rydberg and valence transitions.  
For the latter, we also perform ADC(2)/{\AVTZ} \cite{Tro97,Dre15} calculations using Q-CHEM 5.3/5.4, \cite{Epi21} as well as TD-CAM-B3LYP/{\AVTZ} \cite{Yan04} calculations with GAUSSIAN, \cite{Gaussian16} in 
an effort to estimate the CT character of the valence ESs. Following Ref.~\citenum{Loo21a}, we consider two CT metrics: (i) the electron-hole distance as determined from the ADC(2) transition density matrix, 
\cite{Pla14,Pla14b} and (ii) the CT distance as obtained with CAM-B3LYP applying the well-known Le Bahers' model. \cite{Leb11c,Ada15}   As detailed elsewhere,  \cite{Loo21a} although these two metrics often 
yield rather consistent estimates of the CT strength, there is no definitive definition of CT state. We consider that a given transition has a non-negligible CT character if the electron-hole distance provided by these 
metrics is greater than 1~\AA.

In a second stage, we apply CC3 \cite{Chr95b,Koc95} to estimate the VTEs of selected ESs with three atomic basis sets, namely {\Pop}, {\AVDZ}, and {\AVTZ}. These calculations are carried out with either  CFOUR,  \cite{Mat20} 
and/or DALTON 2018, \cite{dalton} the latter being able to provide CC3 triplet transitions.  For all tested cases, the two codes provide the same VTEs within $\pm$0.001 eV. When the CC3/{\AVTZ} calculations were achievable with 
Dalton, we provide below $\Td$, that is, the percentage of single excitations involved in a given transition at this level of theory. Consistent with earlier works, \cite{Sch08,Loo18a,Loo19c} $\Td$ is typically much larger for 
triplet transitions than for their singlet counterparts, the latter having a larger contributions from double excitations. This is why, for all singlet ESs considered herein, we also report CCSDT  \cite{Nog87,Scu88,Kuc01,Kow01,Kow01b}
excitation energies (computed with CFOUR \cite{Mat20}) obtained with at least one of the two double-$\zeta$ basis sets, the difference between the CC3 and CCSDT VTEs providing a first hint at the convergence of the VTEs with 
respect to the maximum excitation degree of the truncated CC series. Of course, neither CC3 nor CCSDT transition energies can be considered as of FCI quality. Thus, one can certainly wonder what would be the impact of ramping 
up further the truncation degree of the CC expansion. However, given the size of the molecules treated here, such task is very clearly beyond computational reach. Besides, we wish to stress that, for most of the molecules depicted 
in Figure \ref{Fig-1}, the present CC3 and CCSDT VTEs are the first being published. Indeed, as detailed in the next Section, all previous efforts typically consider significantly lower levels of theory.

In a third phase, we assess the performances of many wave function methods using these new TBEs. All these benchmark calculations employ the {\AVTZ} basis set. Consistent with the QUEST database, \cite{Ver21} we test:
CIS(D),  \cite{Hea94,Hea95} EOM-MP2,  \cite{Sta95c}  CC2,  \cite{Chr95,Hat00} CCSD, \cite{Pur82,Scu87,Koc90b,Sta93,Sta93b} STEOM-CCSD, \cite{Noo97,Dut18}  CCSD(T)(a)*, \cite{Mat16} CCSDR(3),  \cite{Chr96b}  CCSDT-3, 
\cite{Wat96,Pro10} CC3,  \cite{Chr95b,Koc95}  ADC(2), \cite{Tro97,Dre15} ADC(3), \cite{Tro02,Har14,Dre15}  and ADC(2.5). \cite{Loo20b} The ADC(2), ADC(3), and EOM-MP2 calculations are performed with Q-CHEM, \cite{Epi21} 
using the resolution-of-the-identity (RI) approximation and tightening the convergence and integral thresholds.  Note that the ADC(2.5) VTEs are the simple average of the ADC(2) and ADC(3) values. 
The CIS(D) and CCSD VTEs are obtained with GAUSSIAN, \cite{Gaussian16} CC2 and CCSDR(3) results are obtained with DALTON \cite{dalton}, whereas CCSD(T)(a)* and CCSDT-3 energies are computed with CFOUR. \cite{Mat20}
STEOM-CCSD calculations are carried out with ORCA 4.2.1 \cite{Nee12} and we report only ESs for which the active character percentage is larger than $98\%$.  In addition, we also evaluate the spin-component scaling (SCS)
and scaled spin-opposite (SOS) CC2 approaches, as implemented in TURBOMOLE 7.3. \cite{Hel08,Turbomole} For ADC(2), we apply two distinct sets of SOS parameters, the ones available in Q-CHEM \cite{Kra13}  and the ones proposed
by TURBOMOLE. \cite{Hel08}
 
\section{Results and discussion}

\subsection{Reference values}

\subsubsection{Azulene}

The ESs of this isomer of naphthalene were investigated in several theoretical works, \cite{Die04,Mur04b,Fal09,Huz12,Pie13,Vos15,Vey20,Loo21a} and we have considered here six singlet ESs (four valence, two Rydberg) and 
four triplet ESs (all valence). We refer the interested reader to Table \ref{Table-1} as well as the SI for further details regarding oscillator strengths and involved MO pairs. The four valence singlet ESs are consistently attributed by experimental
 \cite{Gil78,Hir78,Fuj83,Bla08,Vos15}  and theoretical \cite{Mur04b,Fal09,Huz12,Pie13,Vos15} approaches. For their Rydberg counterparts, the assignments are clearly more challenging (see below). \cite{Bla08,Pie13} 
According to the analysis of the CT strengths presented in our earlier work, \cite{Loo21a} two ESs present a mild CT character with an electron-hole separation around $1$ \AA.  For these two states, the present work is, 
as far as we are aware, the first to present CC results including iterative triples for azulene.

\begin{table*}[htp]
\caption{\small Vertical transition energies (in eV) of azulene. We provide the symmetry of all states as well as the nature of the transition. The rightmost columns list selected data from the literature.}
\label{Table-1}
\footnotesize
\vspace{-0.3 cm}
\begin{tabular}{l|cccc|ccccccc}
\hline
&\multicolumn{2}{c}{\Pop} & \AVDZ & \AVTZ   & \multicolumn{7}{c}{Lit.} \\
State&  CC3 & CCSDT & CC3 & CC3 &  Exp.& Exp. & Exp. & Th. & Th. & Th. & Th.  \\
\hline																
$^1B_2$ (Val, $\pp$)	&2.171	&2.163	&2.177	&2.169	&1.77$^a$&1.72$^b$	&1.77$^e$	&1.96$^f$		&2.25$^g$	&1.62$^h$	&1.83$^i$		\\
$^1A_1$ (CT, $\pp$)	&3.959	&3.965	&3.878	&3.843	&3.57$^a$ &3.56$^c$	&3.56$^e$	&3.81$^f$		&3.99$^g$	&3.41$^h$	&3.46$^i$		\\
$^1B_2$ (CT, $\pp$)	&4.561	&4.580	&4.523	&4.491	& 		&4.22$^d$	&4.23$^e$	&4.15$^f$		&4.66$^g$	&4.08$^h$	&4.13$^i$		\\
$^1A_2$ (Ryd.)		&5.012	&5.031	&4.783	&4.855	& 		&4.40$^d$	&4.72$^e$	&			&4.79$^g$	&			&			\\
$^1A_1$ (Val, $\pp$)	&5.018	&5.060	&4.941	&4.914	& 		&			&4.40$^e$	&4.94$^f$		&5.05$^g$	&4.77$^h$	&4.50$^i$		\\
$^1B_1$ (Ryd.)		&5.338	&5.355	&5.216	&5.285	& 		&			&5.19$^e$	&			&5.22$^g$	&			&			\\	
$^3B_2$ (Val, $\pp$)	&2.211	&		&2.189	&		&1.72$^a$&			&			&			&			&			&1.76$^i$		\\
$^3A_1$ (Val, $\pp$)	&2.430	&		&2.466	&		&2.38$^a$&			&			&			&			&			&2.26$^i$		\\
$^3A_1$ (Val, $\pp$)	&2.923	&		&2.900	&		&2.85$^a$&			&			&			&			&			&2.70$^i$		\\
$^3B_2$ (Val, $\pp$)	&4.203	&		&4.161	&		&3.86$^a$&			&			&			&			&			&3.87$^i$		\\
\hline
\end{tabular}
\vspace{-0.3 cm}
\begin{flushleft}
\begin{footnotesize}
$^a${Photoelectron spectroscopy from Ref.~\citenum{Vos15};}
$^b${0-0 energy in frozen matrix from  Ref.~\citenum{Gil78};}
$^c${0-0 energy from fluorescence study of Ref.~\citenum{Hir78};}
$^d${0-0 energy from the fluorescence spectrum of the jet-cooled derivative in Ref.~\citenum{Fuj83};}
$^e${``Electronic energy'' from pump-probe experiments of Ref.~\citenum{Bla08}. Here, we simply assigned the two lowest Rydberg according to their energetic ordering;}
$^f${CASPT2/6-31G(d) values from Ref.~\citenum{Mur04b};}
$^g${CCSDR(3)/DZ+P values from Ref.~\citenum{Fal09};}
$^h${$\delta$-CR-EOMCC(2,3)/cc-pVDZ values from Ref.~\citenum{Pie13};}
$^i${DFT/MRCI values from Ref.~\citenum{Vos15}, determined at the anionic geometry.}
\end{footnotesize}
\end{flushleft}
\end{table*}

For the singlet transitions, the variations between CCSDT/{\Pop} and CC3/{\Pop} are very mild (roughly $\pm0.02$ eV) except for the second $A_1$ ES for which a slightly larger effect can be noticed ($+0.04$ eV). 
In the same vein, the basis set effects are following the expected trends,  with maximal variations of approximately $0.07$ eV between the {\AVDZ} and {\AVTZ} singlet VTEs at the CC3 level. There is therefore a high 
level of consistency between the various values collected in Table \ref{Table-1}.

There are several measurements of the optical properties of azulene performed with diverse experimental techniques, \cite{Gil78,Hir78,Fuj83,Bla08,Vos15} some of which are summarized at the right- hand-side of Table \ref{Table-1}. 
Let us start with the singlet transitions. Our VTE for the lowest $^1B_2$ is significantly below the measured 0-0 energy, by approximately $-0.4$ eV. We recall that this is actually the expected trend: the vertical energies do not account for the geometric 
relaxation in the ES, nor the difference of zero-point vibrational energies of the two states, and should therefore exceed their 0-0 counterparts. \cite{Die04,San16b}  The same trend pertains for all higher-energy valence transitions with typical shifts going 
from $-0.3$ to $-0.5$ eV between the VTEs listed in Table \ref{Table-1} and the experimental values. For the two Rydberg ESs, attributing the two lowest experimental peaks to the two lowest theoretical Rydberg ESs, one obtains
trends that are not incompatible with the measurements. Yet, one should clearly be very cautious and more experimental and theoretical analyses would be welcome for these particular ESs.  For the triplets transitions, our estimates are slightly 
(for the two $^3A_1$ states) or significantly (for the two $^3B_2$ states) larger than the measurements but provide the same ranking.

As compared to the first CASPT2 values published almost two decades ago, \cite{Mur04b} our CC3/{\AVTZ} values are similar for the $^1A_1$ VTEs, but higher for the $^1B_2$ excitations. If one compares to the previous high-level CC 
estimates including perturbative triples corrections (and performed with relatively compact basis sets), \cite{Fal09,Pie13} one notes that the present energies are slightly lower/higher than the CCSDR(3) values of Ref.~\citenum{Fal09} 
for the valence/Rydberg transitions, whereas, surprisingly, the $\delta$-CR-EOMCC(2,3) results \cite{Fra11a} of Ref.~\citenum{Pie13} are significantly lower for a reason that remains unclear to us. 
Note, however, that the CCSD excitation energies of Ref.~\citenum{Pie13} and ours (see Section \ref{sec:bench}) are in excellent agreement.

\subsubsection{BOD and BTD}

\begin{table*}[htp]
\caption{\small Vertical transition energies (in eV) of BOD and BTD. See caption of Table \ref{Table-1} for more details.}
\label{Table-2}
\footnotesize
\vspace{-0.3 cm}
\begin{tabular}{ll|ccccc|cc}
\hline
\multicolumn{8}{c}{2,1,3-benzoxadiazole (BOD)}\\
&&\multicolumn{2}{c}{\Pop} & \multicolumn{2}{c}{\AVDZ} & \AVTZ  &  \multicolumn{2}{c}{Lit.}\\
State& $\Td$& CC3 & CCSDT & CC3 & CCSDT & CC3 & Exp. & Th.\\
\hline
$^1B_2$ (Val, $\pp$)		&88.6	&4.706	&4.794	&4.575	&4.661	&4.520	&4.00$^a$ 	&4.16$^b$\\
$^1A_1$ (Val, $\pp$)		&83.5	&4.989	&4.990	&4.940	&4.945	&4.906	&4.40$^a$ 	&4.85$^b$\\
$^1A_2$ (Val, $\np$)		&86.9	&5.461	&5.483	&5.368	&5.396	&5.284	\\
$^1B_1$ (Val, $\nsp$)	&85.6	&5.997	&6.009	&5.899	&5.915	&5.833	\\
$^3B_2$ (Val, $\pp$)		&97.5	&2.763	&		&2.751	&		&2.739	\\
$^3A_1$ (Val, $\pp$)		&97.2	&4.181	&		&4.118	&		&4.084	\\			
\hline
\multicolumn{8}{c}{2,1,3-benzothiadiazole (BTD)}\\
&&\multicolumn{2}{c}{\Pop} & \multicolumn{2}{c}{\AVDZ}  & \AVTZ  & \multicolumn{2}{c}{Lit.}\\
State& $\Td$& CC3 & CCSDT &  CC3 & CCSDT  & CC3 & Exp.& Th.\\
\hline
$^1B_2$ (CT,  $\pp$)	&86.1	&4.419	&4.481	&4.301	&4.363	&4.229	&3.77$^a$&3.94$^b$	\\
$^1A_1$ (Val,  $\pp$)	&86.5	&4.465	&4.477	&4.405	&4.417	&4.359	&4.05$^a$&4.11$^b$	\\
$^1A_2$ (Val, $\np$)		&87.7	&4.977	&4.984	&4.886	&4.897	&4.795	&		&			\\
$^1B_1$ (Val, $\nsp$)	&86.1 	&5.616	&5.620	&5.520	&5.525	&5.417	&		&			\\
$^3B_2$ (Val, $\pp$)		&97.3 	&2.833	&		&2.836	&		&2.820	&2.28$^c$&			\\
$^3A_1$ (Val, $\pp$)		&97.3 	&3.646	&		&3.551	&		&3.485	&		&			\\
\hline
\end{tabular}
\vspace{-0.3 cm}
\begin{flushleft}
\begin{footnotesize}
$^a${Vapor phase 0-0 energies from Ref.~\citenum{Hol69};}
$^b${ADC(3)/{\AVDZ} value from Ref.~\citenum{Prl16b}; for the lowest singlet of BTD, a value of $4.28$ eV was reported as basis set extrapolated TBE in Ref.~\citenum{Loo21a}
whereas a $4.15$ eV TD-DFT estimate is given in Ref.~\citenum{Ref11}; }
$^c${0-0 phosphorescence  measured in a frozen dichlorobenzene matrix from Ref.~\citenum{Lin78}. The same work reports a 0-0 energy of $3.52$ eV for the lowest singlet, significantly
redshifted as compared to the vapor measurement of Ref.~\citenum{Hol69}.}
\end{footnotesize}
\end{flushleft}
\end{table*}

Our results for 2,1,3-benzoxadiazole (BOD, also named benzofurazan in the literature) and 2,1,3-benzothiadiazole (BTD) are listed in Table \ref{Table-2}.  These building blocks are popular
in many applications, e.g., they can be used as fluorescent probes for the former \cite{Liu11f} and as an accepting moiety in solar cell materials for the latter.\cite{Li12e}  Unsurprisingly, the ordering of the ESs is the same for the two 
compounds, the sulfur-bearing molecule presenting more redshifted values, except for the lowest triplet state. The lowest ES of BTD has a significant CT character, \cite{Loo21a} 
whereas its BOD counterpart does not display significant separation between the electron and the hole according to popular metrics. \cite{Leb11c,Pla14,Pla14b} As can be seen in Table 
\ref{Table-2}, rather usual basis set effects are obtained with regular decrease of the transition energies as the basis set size increases, although the amplitude of the changes are strongly
state-dependent, as illustrated by the ``insensitivity'' to basis set effect of the lowest triplet state. For the two lowest triplet ESs of $\pp$ nature, very large $\Td$ ($> 97$\%) are calculated, and one 
can be confident that the CC3 values are accurate. For the singlet ESs, the differences between the CC3 and CCSDT results are of the order of  $0.02$ eV, except for the lowest $^1B_2$ ES:
the CC3 estimates ($4.71$ and $4.42$ eV for BOD and BTD, respectively) are significantly smaller than their CCSDT counterparts ($4.79$ and $4.48$ eV, respectively). Such trend is typical of CT states, \cite{Koz20,Loo21a}
and seems to apply to the first ES of BOD though the tested metrics did not revealed a significant CT character.

For substituted furazans, several TD-DFT studies can be found, \cite{Tsu09,Bro12,Chi14c} but apparently the only previous investigation of the building block itself is the work of Prlj \emph{et al.}
\cite{Prl16b} who studied the $L_a$/$L_b$ (or $^1B_2$/$^1A_1$) ordering in several bicyclic systems. This work reports CC2 transition energies closely matching the present values (4.588 and 4.887 eV), whereas
the ADC(3) VTEs show a larger gap between the two ESs (see rightmost column in Table \ref{Table-2}). The vapour spectrum of BOD was measured in 1969, \cite{Hol69} and the experimental 
0-0 energies are shifted by approximately $-0.5$ eV compared to our VTEs, which is a typical trend.  For BTD, more measurements are available \cite{Hol69,Gor71,Hen75,Lin78} and the lowest transitions were treated 
previously with TD-DFT \cite{Ref11,Chi14b,Men15,Prl16b} and wave function approaches, \cite{Prl16b,Loo21a} a few relevant values being summarized in Table \ref{Table-2}.  The gap between the two lowest 
singlet transitions is around $0.10$ eV according to our data, the $^1B_2$ ES being the lowest-energy state. Previous CC2 and ADC(3) values report a similar pattern. \cite{Prl16b} In contrast CCSD/{\AVTZ} yields almost degenerated 
transitions but with the incorrect ordering (see the SI). The experimental BTD 0-0 energies of Ref.~\citenum{Hol69} are red-shifted by $-0.23$ and $-0.35$ eV as compared to those of BOD for the $^1B_2$ and $^1A_1$ 
 transitions, respectively. The CC3/{\AVTZ} heteroatomic shifts of the vertical energies are of the same order of magnitude, i.e., $-0.29$ and $-0.55$ eV.

\subsubsection{DPP}

Due to its intense absorption around 500 nm, DPP is also an extraordinary popular moiety to design dyes used in automotive paints, light harvesting applications, or fluorescent sensors. \cite{Grz15}
While there exist many TD-DFT calculations of DPP-containing compounds in the literature, we could not find previous theoretical works devoted to the chromogen itself, but for a 2009
TD-DFT contribution, \cite{Lun09b} and studies limited to the solvation effects of the lowest transition. \cite{Chi14b,Men15}  According to our calculations (Table \ref{Table-3}) the lowest 
singlet ES is a bright $^1B_u$ that interestingly lies more than $1.5$ eV above the corresponding triplet, hinting at small intersystem crossing.  Next, one finds two (nearly) dark  ESs of $^1A_u$ ($\np$) and 
$^1A_g$ ($\pp$) symmetries, whereas the fourth ESs is a dark $^1B_g$ ($\pp$). For all eight transitions listed in Table \ref{Table-3}, the basis set effects are rather small, with variations of ca.~$-0.05$ to 
$-0.10$ eV between the {\Pop} and {\AVTZ} VTEs. With the former basis set, we could perform CCSDT calculations, and the outcome hints that the CC3 VTEs are slightly too low by approximately $0.03$-$0.04$ eV 
for the four singlet ESs. As in BOD and BTD, very large $\Td$ are found for the triplet ESs, and one can likely view the CC3/{\AVTZ} excitation energies as reliable TBEs for the triplets.

\begin{table*}[htp]
\caption{\small Vertical transition energies (in eV) of DPP. See caption of Table \ref{Table-1} for more details.}
\label{Table-3}
\footnotesize
\vspace{-0.3 cm}
\begin{tabular}{ll|cccc|c}
\hline
&&\multicolumn{2}{c}{\Pop} & \AVDZ & \AVTZ  & Lit. \\
State& $\Td$& CC3 & CCSDT & CC3 & CC3 & Th. \\
\hline
$^1B_u$ (Val, $\pp$)	&88.4	&3.650	&3.683	&3.541	&3.535&	3.42$^a$\\
$^1A_u$ (Val, $\np$)	&83.7	&3.953	&3.989	&3.907	&3.863&	3.75$^a$\\
$^1A_g$ (Val, $\pp$)	&87.0	&3.959	&4.009	&3.872	&3.910&	3.95$^a$\\
$^1B_g$ (Val, $\np$)	&81.5 	&4.397	&4.426	&4.324	&4.309&	4.21$^a$\\
$^3B_u$ (Val, $\pp$)	&97.4	&1.957	&		&1.923	&1.927&	\\
$^3A_g$ (Val, $\pp$)	&97.3	&3.804	&		&3.751	&3.743&	\\
$^3A_u$ (Val, $\np$)	&94.9	&3.867	&		&3.787	&3.781&	\\
$^3B_g$ (Val, $\np$)	&94.5 	&4.310	&		&4.240	&4.226&	\\
\hline
\end{tabular}
\vspace{-0.3 cm}
\begin{flushleft}
\begin{footnotesize}
$^a${TD-PBE0/6-311++G(d,p)  from Ref.~\citenum{Lun09b}.}
\end{footnotesize}
\end{flushleft}
\end{table*}

\subsubsection{FF, PP, and TT}

 Furo[3,2-$b$]furan (FF), 1,4-dihydropyrrolo[3,2-$b$]pyrrole (PP) and thieno[3,2-$b$]thiophene (TT) are centrosymmetric bicyclic molecules encompassing two identical and fused five member cycles.  TT is often used as
 a linker in $\pi$-delocalized polymers, \cite{Mcc09} whereas PP is an increasingly popular accepting moiety, notably in quadrupolar systems showing large nonlinear optical responses. \cite{Tas19} Our
 results are collected in Table \ref{Table-4}.

According to our calculations, the lowest  singlet ES of FF is of  Rydberg character. It presents almost the same VTE as the bright $^1B_u$, the latter being typical in $\pi$-delocalized dyes of $C_{2h}$ symmetry. 
The hallmark dark $\pp$  $^1A_g$ ES lies approximately $0.6$ eV higher. As expected it has a smaller single excitation character ($\Td=82.6$\%) than the other ESs treated here, but the difference between the 
CC3 and CCSDT VTEs remains small ($-0.02$ eV). This can be compared to the $^1A_g$ ES in hexatriene ($\Td=65.3$\%) for which the CC3-CCSDT difference is about five times larger. \cite{Ver21}  
We have also found two other Rydberg transitions of $B_g$ symmetry, and two $\pp$ triplet ESs with very large $\Td$, strongly redshifted compared to the corresponding singlet excitations. Going 
from {\AVDZ} to {\AVTZ} induces an increase/decrease of the VTEs for the Rydberg/valence transitions. In PP, the CC3/{\AVTZ} calculations indicate that the four lowest singlet transitions are of Rydberg nature 
(the $\pp$ $^1B_u$ lies higher, at 5.499 eV according to CC3/{\AVTZ}).  All these four ESs have a strong single-excitation character. In contrast, in the triplet manifold, the lowest-energy ES has a valence character, the two following ones being 
of Rydberg nature. The picture is vastly different in TT, in which one first notices two low-lying nearly-degenerated singlet valence ESs of $B_u$ symmetry (see also the discussion below), the Rydberg 
transitions appearing at slightly higher energy. In the triplet manifold of TT, three valence $\pp$ ESs could be identified. All seven ESs of TT listed in Table \ref{Table-4} are characterized by $\Td > 85$\%, 
highly similar CC3 and CCSDT VTEs, and rather mild basis set effects.

\begin{table*}[htp]
\caption{\small Vertical transition energies (in eV) of FF, PP, and TT. See caption of Table \ref{Table-1} for more details.}
\label{Table-4}
\footnotesize
\vspace{-0.3 cm}
\begin{tabular}{ll|ccccc|c}
\hline
\multicolumn{8}{c}{Furo[3,2-$b$]furan (FF)}\\
&&\multicolumn{2}{c}{\Pop} &  \multicolumn{2}{c}{\AVDZ}  & \AVTZ  & Lit.\\
State& $\Td$& CC3 & CCSDT & CC3 & CCSDT & CC3 & Th.\\
\hline
$^1A_u$ (Ryd)		&93.4	&5.590	&5.602	&5.357	&5.361	&5.430&	\\
$^1B_u$ (Val, $\pp$)	&91.5	&5.644	&5.672	&5.500	&5.526	&5.463&5.63$^a$	\\
$^1B_g$ (Ryd)		&93.4	&5.985	&6.002	&5.783	&5.789	&5.859&	\\
$^1B_g$ (Ryd)		&93.1	&6.148	&6.165	&5.934	&5.942	&5.993&	\\
$^1A_g$ (Val, $\pp$)	&82.6	&6.250	&6.228	&6.080	&6.067	&6.040&	\\
$^3B_u$ (Val, $\pp$)	&97.9	&3.661	&		&3.601	&		&3.578&	\\
$^3A_g$ (Val, $\pp$)	&98.2	&4.956	&		&4.897	&		&4.869&	\\
\hline
\multicolumn{8}{c}{1,4-Dihydropyrrolo[3,2-$b$]pyrrole (PP)}\\
&&\multicolumn{2}{c}{\Pop} & \multicolumn{2}{c}{\AVDZ} & \AVTZ  \\
State& $\Td$& CC3 & CCSDT & CC3 & CCSDT & CC3 \\
\hline
$^1A_u$ (Ryd)		&92.8	&4.558	&4.563	&4.454	&4.445	&4.545\\
$^1B_g$ (Ryd)		&92.5	&4.761	&4.768	&4.656	&4.649	&4.746\\
$^1A_u$ (Ryd)		&92.0	&5.088	&5.078	&5.020	&4.994	&5.133\\
$^1B_g$ (Ryd)		&93.1	&5.275	&5.281	&5.067	&5.055	&5.145\\
$^3B_u$ (Val, $\pp$)	&97.9	&3.921	&		&3.867	&		&3.841\\
$^3A_u$ (Ryd)		&97.4	&4.529	&		&4.431	&		&4.524\\
$^3B_g$ (Ryd)		&97.3	&4.739	&		&4.641	&		&4.733\\
\hline 
\multicolumn{8}{c}{Thieno[3,2-$b$]thiophene (TT)}\\
&&\multicolumn{2}{c}{\Pop} &  \multicolumn{2}{c}{\AVDZ}  & \AVTZ & Lit. \\
State& $\Td$& CC3 & CCSDT & CC3 & CCSDT & CC3 & Th. \\
\hline
$^1B_u$ (Val, $\pp$)	&87.5	&5.100	&5.096	&5.003	&5.004	&4.964&5.04$^b$\\
$^1B_u$ (Val, $\pp$)	&90.6	&5.439	&5.460	&5.301	&5.320	&5.224&5.29$^b$\\	
$^1B_g$ (Ryd)		&90.3	&5.465	&5.465	&5.455	&5.454	&5.412&\\
$^1A_u$ (Ryd)		&91.6	&5.487	&5.485	&5.483	&5.474	&5.518&\\
$^3B_u$ (Val, $\pp$)	&97.7	&3.488	&		&3.490	&		&3.465&\\
$^3B_u$ (Val, $\pp$)	&97.2	&4.376	&		&4.301	&		&4.261&\\
$^3A_g$ (Val, $\pp$)	&97.9	&4.669	&		&4.613	&		&4.580&\\
\hline
\end{tabular}
\vspace{-0.3 cm}
\begin{flushleft}
\begin{footnotesize}
$^a${ADC(2)/cc-pVQZ value from Ref.~\citenum{Prl15};}
$^b${SAC-CI/cc-pVTZ value from Ref.~\citenum{Prl15}.}
\end{footnotesize}
\end{flushleft}
\end{table*}

For FF, we have found only one previous study, \cite{Prl15} that reported an ADC(2) value for the lowest $^1B_u$ transition in reasonable agreement with the present estimate.  For the non-substituted PP
we could not find previous theoretical nor experimental works. In contrast, TT has been studied using various levels of theory. \cite{Chi14b,Prl15,Men15} For its two lowest ESs, a refined theoretical study  
 proposed by the Corminboeuf group, \cite{Prl15} highlighted the challenge of obtaining an accurate ordering with TD-DFT (almost) independently of the selected exchange-correlation functional.  At the CCSD/{\AVTZ}
 level, these two ESs are quite strongly mixed in terms of underlying MOs (see Table S10 in the SI).  Our best estimates actually provide the same ordering as the CC2 and SAC-CI approaches used in Ref.~\citenum{Prl15}, i.e.,
the lowest ES is dominated by a $\text{HOMO}-1$ to LUMO character, whereas the second ES is mainly corresponding to a HOMO to LUMO excitation.

\subsubsection{Phthalazine and quinoxaline}

We also consider two diazanaphthalenes having to $C_{2v}$ point group symmetry that are popular building blocks in dye chemistry, namely phthalazine and quinoxaline. \cite{Wu13c,Thi16} The latter presents lowest singlet and triplet 
ESs of different chemical natures, making quinoxaline derivatives particularly appealing for purely organic TADF applications, thanks to an exceptionally efficient intersystem crossing. \cite{Shi15b} We report in Table \ref{Table-5} 
numerous singlet and a few triplet excited states for these two compounds.

\begin{table*}[htp]
\caption{\small Vertical transition energies (in eV) of phthalazine and quinoxaline. See caption of Table \ref{Table-1} for more details.}
\label{Table-5}
\footnotesize
\vspace{-0.3 cm}
\begin{tabular}{l|cccc|ccccc}
\hline
\multicolumn{10}{c}{Phthalazine}\\
&\multicolumn{2}{c}{\Pop} & \AVDZ & \AVTZ   & \multicolumn{4}{c}{Lit.} \\
State&  CC3 & CCSDT & CC3 & CC3 &  Exp.& Exp. &  Th. & Th.  \\
\hline	
$^1A_2$ (CT, $\np$)	&3.986	&4.012	&3.889	&3.872	&3.61$^a$	&3.01$^c$	&3.68$^d$	&3.74$^e$\\
$^1B_1$ (CT, $\np$)	&4.427	&4.446	&4.317	&4.283	&3.92$^a$	&3.72$^c$	&4.12$^d$	&4.20$^e$\\
$^1A_1$ (Val, $\pp$)	&4.539	&4.517	&4.501	&4.473	&4.13$^a$	&4.09$^c$	&			&4.46$^e$\\
$^1B_2$ (Val, $\pp$)	&5.364	&5.406	&5.201	&5.146	&4.86$^a$	&4.59$^c$	&5.27$^d$	&4.98$^e$\\
$^1B_1$ (CT, $\np$)	&5.662	&5.690	&5.561	&5.520	&			&			&			&\\
$^1A_2$ (Mixed)	&5.954	&6.052	&5.745	&5.744	&			&			&			&\\
$^1A_2$ (CT, $\np$)	&6.037	&6.043	&5.930	&5.870	&			&5.33$^c$	&			&\\
$^1A_1$ (Val, $\pp$)	&6.243	&6.204	&6.185	&6.148	&			&			&			&\\
$^1A_2$ (Ryd)		&6.606	&6.620	&6.362	&6.430	&			&			&			&\\
$^1B_2$ (Ryd)		&6.379	&6.408	&6.133	&6.234	&			&			&			&\\
$^1A_1$ (Val, $\pp$)	&6.510	&6.554	&6.410	&6.364	&5.84$^a$	&			&			&\\
$^3B_2$ (Val, $\pp$)	&3.452	&		&3.440	&3.430	&2.85$^b$	&2.74$^c$	&			&3.42$^e$\\
$^3A_2$ (CT, $\np$)	&3.721	&		&3.628	&3.626	&			&			&			&3.48$^e$\\
$^3B_1$ (CT, $\np$)	&3.801	&		&3.720	&3.711	&			&			&			&3.67$^e$\\
$^3A_1$ (Val, $\pp$)	&4.333	&		&4.266	&4.224	&			&			&			&4.30$^e$\\
\hline
\multicolumn{10}{c}{Quinoxaline}\\
&\multicolumn{2}{c}{\Pop} & \AVDZ & \AVTZ   & \multicolumn{5}{c}{Lit.} \\
State&  CC3 & CCSDT & CC3 & CC3 &  Exp.& Exp. & Exp. & Th. & Th.  \\
\hline	
$^1B_1$ (Val, $\np$)	&3.944	&3.954	&3.831	&3.790	&3.58$^a$	&3.36$^c$&3.36$^f$	&3.83$^d$		&3.76$^e$	\\
$^1A_1$ (Val, $\pp$)	&4.333	&4.318	&4.295	&4.263	&4.00$^a$	&3.97$^c$&3.96$^f$	&			&4.26$^e$	\\
$^1B_2$ (CT, $\pp$)	&4.772	&4.827	&4.643	&4.586	&4.34$^a$	&4.09$^c$&		&4.89$^d$	&4.45$^e$	\\
$^1A_2$ (Val, $\np$)	&5.220	&5.233	&5.109	&5.087	&			&		&		&5.39$^d$	&5.03$^e$	\\
$^1A_2$ (Val, $\np$)	&5.549	&5.547	&5.442	&5.390	&			&		&		&			&	\\
$^1A_1$ (CT, $\pp$)	&5.770	&5.761	&5.716	&5.674	&5.35$^a$	&5.33$^c$&5.36$^f$	&			&	\\%
$^1B_1$ (CT, $\np$)	&6.368	&6.433	&6.163	&6.140	&			&5.70$^c$&		&			&	\\
$^1B_2$ (Val, $\pp$)	&6.489	&6.511	&6.332	&6.277	&			&		&		&			&	\\
$^3B_2$ (Val, $\pp$)	&3.286	&		&3.270	&3.255	&			&2.68$^c$&		&			&3.26$^e$	\\%
$^3B_1$ (Val, $\np$)	&3.461	&		&3.368	&3.352	&			&3.04$^c$&		&			&3.31$^e$	\\%
$^3A_1$ (Val, $\pp$)	&4.012	&		&3.919	&3.875	&			&		&		&			&4.76$^e$	\\%
\hline
\end{tabular}
\vspace{-0.3 cm} 
\begin{flushleft}
\begin{footnotesize}
$^a${MCD spectra measured in $n$-heptane from Ref.~\citenum{Kai78};}
$^b${0-0 phosphoprescence in a frozen matrix from Ref.~\citenum{Lim70};}
$^c${0-0 energy collected in Ref.~\citenum{Inn88} (see references therein);}
$^d${CASPT2/cc-pVQZ values from Ref.~\citenum{Mor09};}
$^e${CC2/{\AVDZ} data from Ref.~\citenum{Eti17}:}
$^f${Gas-Phase 0-0 energies from Ref.~\citenum{Gla70}}
\end{footnotesize}
\end{flushleft}
\end{table*}

The impact of including full (CCSDT) rather than approximate (CC3) triples is never large yet appears not insignificant for several ESs of phthalazine with increase of the VTEs by $0.04$ eV for the two highest ESs considered in Table \ref{Table-5}
as well as for the lowest singlet state of $B_2$ symmetry.  For quinoxaline, the largest differences between the CCSDT/{\Pop} and CC3/{\Pop} transition energies are found for two CT transitions: $^1B_2$ ($+0.06$ eV) and 
$^1B_1$ ($+0.07$ eV). As already explained, such slight CC3 underestimation of the CT VTEs has been reported before. \cite{Koz20,Loo21a} Regarding basis set effects, nothing beyond expectations emerges with a mean 
decrease of $-0.04$ eV when switching from {\AVDZ} to {\AVTZ} for quinoxaline, the actual shift being not very ES-dependent (the corrections go from $-0.015$ to $-0.057$ eV). The scenario is similar for phthalazine CT and valence excitations 
(mean impact of going from {\AVDZ} to {\AVTZ} is $-0.03$ eV, with values ranging from $-0.001$ to $-0.060$ eV), but both the sign and magnitude of the basis corrections are vastly different for the two Rydberg transitions.

Several theoretical and experimental transition energies have been previously given for both systems, \cite{Lim70,Gla70,Kai78,Inn88,Mor09,Eti17} though, to the best of our knowledge, none reported so many ESs, nor considered third-order CC models.
We can thus consider the VTEs of Table \ref{Table-5} as the most accurate published to date. When comparing to experimental data, it appears that, as for the above-discussed compounds,  the VTEs exceed the energies of the 
experimental peaks \cite{Kai78} typically by $0.2$-$0.4$ eV, whereas even larger differences are found with respect to the measured 0-0 energies. \cite{Inn88}  The six CASPT2/cc-pVQZ values of Ref.~\citenum{Mor09} are in 
reasonable agreement with the present CC3/{\AVTZ} results with a mean absolute deviation (MAD) of $0.25$ eV, the usual CASPT2 trend of yielding too small VTEs being clearly highlighted here, except for the first $^1B_2$ transition in phthalazine. 
The deviations with respect to the CC2/{\AVDZ}  data of Ref.~\citenum{Eti17} are small (MAD of $0.09$ eV) if one excludes the $^3A_1$ transition of quinoxaline for which a very large difference of $0.89$ eV is found between the present value 
and the literature estimate. Given that the $\Td$ value associated with this state is very large (96.9\%) and that our CC2/{\AVTZ} result computed on the geometry used in the present study is only $0.16$ eV off its CC3 counterpart, this large 
difference is somewhat surprising.

\subsubsection{TTF}

Tetrathiafulvalene (TTF) possesses very specific redox properties, \cite{Nie00,Sai07}  but its optical signatures are also of interest, notably due to its accepting properties in CT complexes \cite{Nie00} as well as to the presence of several low-lying 
$\ps$ transitions, that are rather unusual in organic derivatives. \cite{Pou02b} We consider here the closed-shell singlet form of TTF (Table \ref{Table-6}), and determine VTEs  for 8 singlet and 6 triplet ESs. It is noteworthy that
we enforce planarity during the optimization, to benefit from the high $D_{2h}$ symmetry, whereas the actual structure is very slightly bent. \cite{Har94}

\begin{table*}[htp]
\caption{\small Vertical transition energies (in eV) of TTF. See caption of Table \ref{Table-1} for more details.}
\label{Table-6}
\footnotesize
\vspace{-0.3 cm}
\begin{tabular}{ll|cccc|ccccc}
\hline
&&\multicolumn{2}{c}{\Pop} & \AVDZ & \AVTZ   & \multicolumn{5}{c}{Litt.}\\
State& $\Td$& CC3 & CCSDT & CC3 & CC3 &  Exp.& Th. & Th. & Th. & Th. \\
\hline
$^1B_{3u}$ (Val, $\ps$)	&90.1	&2.958	&2.971	&2.866	&2.787	&2.72$^a$	&2.08$^b$&2.57$^c$&2.68$^d$&2.42$^e$	\\
$^1B_{2u}$ (Val, $\pp$)	&87.9	&4.159	&4.190	&3.795	&3.742	&3.35$^a$	&3.05$^b$&3.30$^c$&3.46$^d$&3.31$^e$	\\
$^1B_{1g}$ (Val, $\ps$)	&90.5	&4.059	&4.067	&4.013	&3.979	&			&3.31$^b$&3.72$^c$&	\\
$^1B_{2g}$ (Val, $\ps$)	&89.3	&4.083	&4.082	&4.057	&4.048	&			&3.41$^b$&3.92$^c$&	\\
$^1B_{3u}$ (Ryd.)		&92.8	&4.192	&4.193	&4.048	&4.113	&			&3.63$^b$&4.25$^c$&	\\
$^1B_{3g}$ (Val, $\pp$)	&86.2	&4.496	&4.521	&4.264	&4.218	&			&3.44$^b$&3.72$^c$&	\\
$^1B_{1u}$ (Val, $\pp$)	&90.6	&4.805	&4.816	&4.601	&4.514	&3.91$^a$	&4.01$^b$&4.28$^c$&4.25$^d$&4.18$^e$	\\
$^1B_{2g}$ (Ryd.)		&92.0	&4.648	&4.649	&4.495	&4.551	&			&3.80$^b$&		&	\\
$^3B_{3u}$ (Val, $\ps$)	&96.7	&2.804	&		&2.727	&2.652	&			&1.92$^b$&2.33$^c$&	\\
$^3B_{1u}$ (Val, $\pp$)	&97.7	&3.158	&		&3.055	&2.993	&			&2.76$^b$&2.64$^c$&	\\
$^3B_{2u}$ (Val,$\pp$)	&97.1	&3.383	&		&3.160	&3.123	&			&2.89$^b$&2.64$^c$&	\\
$^3B_{3g}$ (Val, $\pp$)	&97.4	&3.565	&		&3.431	&3.400	&			&3.11$^b$&2.90$^c$&	\\
$^3B_{1g}$ (Val, $\ps$)	&96.8	&3.862	&		&3.833	&3.803	&			&		&		&	\\
$^3B_{2g}$ (Val, $\ps$)	&96.8	&3.946	&		&3.942	&3.916	&			&		&		&\\
\hline
\end{tabular}
\vspace{-0.3 cm}
\begin{flushleft}
\begin{footnotesize}
$^a${$\lambda_\mathrm{max}$ in hexane from Ref.~\citenum{Eng77}; very similar values have been reported in other measurements;\cite{Cof71,Wul77}}
$^b${MS-CASPT2 values from Ref.~\citenum{Pou02b};}
$^c${TD-DFT (B3P86/{\AVDZ}) values from Ref.~\citenum{Pou02};}
$^d${CC2/aug-TZVPP values from Ref.~\citenum{Ker09};}
$^e${$\pi$-CASPT2C/aV(T+d)Z values from Ref.~\citenum{Ker09};}
\end{footnotesize}
\end{flushleft}
\end{table*}

The basis set effects seem under control for TTF. We observe, in particular, limited changes between the {\AVDZ} and {\AVTZ} data, with an average downshift of $-0.03$ eV (MAD: $0.05$ eV), and a maximal change 
of $-0.09$ eV ($^1B_{1u}$). The two Rydberg ESs are the only ones for which going from the double- to the triple-$\zeta$ basis set actually increases the computed transition energy, a common pattern for
the present set of molecules.  It is also noteworthy that for all transitions, the differences between the CCSDT and CC3 transitions, as given with the {\Pop} basis set, are small, none of them exceeding $0.03$ eV, consistent
with the fact  that all $\Td$ are rather large. All these elements give us confidence that the reported VTEs are accurate.

Interestingly, previous TD-DFT, \cite{Pou02,Ker09} CC2, \cite{Ker09} and CASPT2 \cite{Pou02b,Ker09} investigations are available for TTF.  If one disregards the Rydberg transitions, the ordering of the CASPT2
ESs reported in 2002 \cite{Pou02b} matches the present CC3 one, but these ``old'' CASPT2 VTEs are notably too low, with underestimations of roughly $-0.4$ eV or more for most states.  The more recent CASPT2 energies
are closer to our present values, yet remain too small. In contrast the TD-DFT calculations performed with the B3P86 global hybrid, \cite{Pou02} as well as CC2 calculations \cite{Ker09} provide estimates closer 
to the current results. The UV/Vis spectra of TTF has been measured by several groups in apolar solvents, \cite{Cof71,Wul77,Eng77} allowing comparisons for three transitions. As expected in such VTE \emph{versus}
$\lambda_{\mathrm{max}}$ comparison, the CC3 values are larger by $0.1$-$0.6$ eV, which indicates that the data of Table \ref{Table-6} are reasonable.

\subsection{Benchmarks}
\label{sec:bench}

To define our TBEs, we typically consider: (i) CCSDT/double-$\zeta$ [6-31+G(d,) or {\AVDZ}] VTEs corrected for basis set effects thanks to the difference between CC3/double-$\zeta$ and CC3/{\AVTZ} for singlet ESs, and (ii) 
select the CC3/{\AVTZ} VTEs for triplet ESs. The complete list of TBEs and each specific protocol employed to obtain these are available in Table S11 in the SI.

CC3 nor CCSDT are exact theories, so before assessing other methods, let us briefly discuss the expected accuracy of the present TBEs.  For the singlet ESs, our reference values are essentially of CCSDT quality and none of the ES 
treated here has a significant double-excitation character. More specifically, all $\Td$ (that could be obtained) exceed 80\%, and the variations between CC3 and CCSDT are typically very small. While this provides confidence that the TBEs 
are accurate, they are likely not precise enough to fairly compare methods that yield highly similar results and are closely related, e.g., CCSDR(3) and CCSD(T)(a)*. For the triplet ESs, our reference values, of CC3/{\AVTZ} quality, are 
likely sufficient to assess the tested methods.  They are three reasons for this assertion. First, all triplet ESs treated here have very large $\Td$. Second, in the full QUEST database, the mean absolute error obtained with CC3/{\AVTZ} 
(as compared to higher levels of theory) is as small as $0.01$ eV. \cite{Ver21} Third, the three other CC methods including corrections for the triples, namely, CCSDR(3),  CCSD(T)(a)*, and CCSDT-3, have not been implemented for triplet 
ESs, so that only methods significantly less advanced than CC3 are benchmarked below.

\begin{figure*}
\centering
 \includegraphics[width=\linewidth]{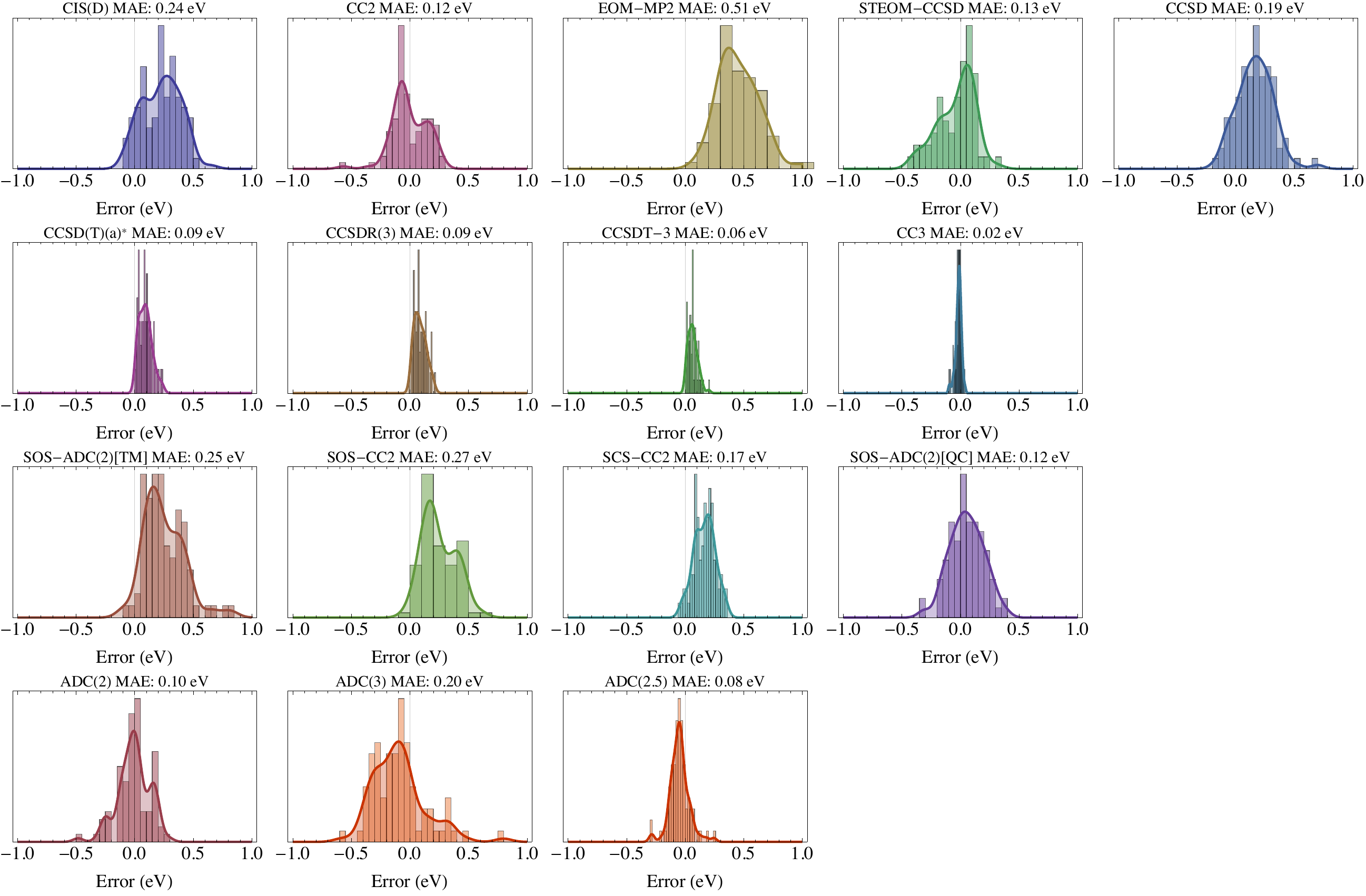}
  \caption{Distribution of the error (in eV) in VTEs (with respect to the TBE/{\AVTZ} values) for various methods. See Table \ref{Table-7} for the values of the corresponding statistical quantities. QC and TM indicate that Q-CHEM and TURBOMOLE scaling factors are considered, respectively. The SOS-CC2 and SCS-CC2 approaches are obtained with the latter code}
   \label{Fig-2}
\end{figure*}

In Table \ref{Table-7}, we report the statistical results obtained considering the full set of data (only singlets are included for all CC methods including triples): mean signed error (MSE), mean absolute error (MAE), root mean
square error (RMSE) and the standard deviation of the errors (SDE).  For the present set of excitations, the distribution of the errors in VTEs (with respect to the TBEs/{\AVTZ}) is represented in Figure \ref{Fig-2}.

\begin{table*}[htp]
\footnotesize
\vspace{-0.3 cm}
\caption{Statistical analysis, taking as reference the TBE/{\AVTZ} values, for various theoretical models.   TM and QC stand for the TURBOMOLE and Q-CHEM
definitions of the scaling factors, respectively. On the right-hand side, we compare the MAE obtained here with the previously values obtained  for smaller
molecules containing 1 to 3 (and 4 to 6) non-hydrogen atoms. \cite{Ver21} All values are in eV. }
\label{Table-7}
\begin{tabular}{lccccc|cccccccc|ccc}
\hline
				&\multicolumn{5}{c}{All states}					&\multicolumn{8}{c}{MAE (subgroup)}		&\multicolumn{3}{c}{MAE (size)}	\\
Method 			& Count 	& MSE 	&MAE 	&RMSE 	&SDE 	&Sing.  	& Trip. 	& Val.	& CT 	& Ryd	&$\np$	&$\pp$	&$\ps$	&1--3  at.	& 4--6  at. 	&This set\\
\hline
CIS(D)			&88		&0.23	&0.24	&0.28	&0.16	&0.20	&0.30	&0.27	&0.28	&0.07	&0.28	&0.28	&0.14	&0.23	&0.22	&0.24	\\
EOM-MP2			&91		&0.50	&0.50	&0.60	&0.34	&0.51	&0.48	&0.51	&0.60	&0.36	&0.61	&0.54	&0.54	&0.13	&0.24	&0.50	\\
STEOM-CCSD		&63		&-0.04	&0.13	&0.16	&0.13	&0.09	&0.17	&0.14	&0.06	&0.12	&0.07	&0.18	&0.18	&0.10	&0.12	&0.13	\\
CC2				&91		&0.00	&0.12	&0.14	&0.15	&0.11	&0.13	&0.12	&0.11	&0.12	&0.09	&0.13	&0.13	&0.19	&0.15	&0.12	\\
SOS-CC2 [TM]		&90		&0.24	&0.24	&0.28	&0.14	&0.24	&0.24	&0.25	&0.27	&0.17	&0.41	&0.17	&0.17	&0.19	&0.23	&0.24	\\
SCS-CC2 [TM]		&91		&0.16	&0.16	&0.19	&0.09	&0.15	&0.19	&0.18	&0.17	&0.08	&0.26	&0.15	&0.15	&0.18	&0.17	&0.16	\\
ADC(2)			&90		&-0.01	&0.10	&0.14	&0.13	&0.10	&0.11	&0.11	&0.10	&0.07	&0.12	&0.12	&0.12	&0.19	&0.14	&0.10	\\
SOS-ADC(2) [TM]	&91		&0.24	&0.25	&0.30	&0.17	&0.28	&0.20	&0.23	&0.25	&0.30	&0.36	&0.15	&0.15	&0.18	&0.21	&0.25	\\
SOS-ADC(2) [QC]	&91		&0.05	&0.12	&0.15	&0.15	&0.14	&0.09	&0.11	&0.18	&0.11	&0.14	&0.10	&0.10	&0.14	&0.12	&0.12	\\
CCSD			&91		&0.16	&0.18	&0.22	&0.15	&0.23	&0.10	&0.17	&0.29	&0.10	&0.31	&0.15	&0.15	&0.07	&0.13	&0.18	\\
ADC(3)			&90		&-0.10	&0.20	&0.24	&0.22	&0.17	&0.24	&0.21	&0.22	&0.10	&0.21	&0.23	&0.23	&0.24	&0.21	&0.20	\\
ADC(2.5)			&88		&-0.05	&0.07	&0.09	&0.08	&0.08	&0.07	&0.07	&0.09	&0.06	&0.06	&0.09	&0.09	&0.10	&0.08	&0.07	\\
CCSD(T)(a)*		&58		&0.09	&0.09	&0.11	&0.06	&0.09	&		&0.10	&0.12	&0.05	&0.14	&0.09	&0.09	&0.03	&0.05	&0.09	\\
CCSDR(3)		&58		&0.09	&0.09	&0.10	&0.06	&0.09	&		&0.10	&0.11	&0.05	&0.13	&0.08	&0.08	&0.03	&0.05	&0.09	\\
CCSDT-3			&58		&0.06	&0.06	&0.08	&0.05	&0.06	&		&0.07	&0.08	&0.04	&0.10	&0.06	&0.06	&0.03	&0.05	&0.06	\\
CC3				&58		&-0.02	&0.02	&0.03	&0.02	&0.02	&		&0.02	&0.03	&0.01	&0.02	&0.03	&0.03	&0.02	&0.01	&0.02	\\
\hline
 \end{tabular}
 \end{table*}

On the left-hand-side of Table \ref{Table-7}, one can find the statistical results obtained considering all ESs. EOM-MP2 appears to strongly overestimate the VTEs and exhibits a large dispersion. In short, it cannot be recommended for 
the present bicyclic compounds. Amongst the computationally light models, CIS(D) is clearly a better option though the associated MAE remains quite large, $0.24$ eV, a rather typical error bar for this level of theory. \cite{Goe10a,Jac15b,Loo18a,Loo20a,Ver21} 
EOM-MP2 and CIS(D) performances are clearly inferior to the ones of two other popular models, namely ADC(2) and CC2 that both deliver very reasonable estimates of the VTEs for the present molecular set, especially ADC(2) that
enjoys a trifling MSE and a MAE as small as $0.10$ eV.  For both CC2 and ADC(2), these trends are compatible with the conclusions obtained from the QUEST database, \cite{Ver21} as well as in many previous works benchmarking ADC(2) 
and/or CC2 for optical properties. \cite{Hat05c,Jac18a,Sch08,Sil10c,Win13,Har14,Jac15b,Kan17,Loo18a,Loo20a} For CC2, the two spin-scaled approaches (SOS and SCS) significantly deteriorate the CC2 MSE and MAE, leading to a clear overestimation.
However, both SOS-CC2 and SCS-CC2 do decrease CC2's dispersion, especially SCS-CC2 that returns a SDE of $0.09$ eV only, a level of consistency that can only be further improved with higher-level approaches according to the data of Table \ref{Table-7}.
Such accuracy drop and gain in consistency of the spin-scaled CC2 methods has been reported before. \cite{Goe10a,Jac15b,Taj20a,Loo20d} For ADC(2), none of the two tested SOS parameters is beneficial as compared to the original
method, but it is very clear that the default Q-CHEM parametrization \cite{Kra13} is superior to its TURBOMOLE analog for the present set. It also appears that STEOM-CCSD is a valuable approach with small MSE, MAE, and SDE. 
However, as explained in the computational details, several ESs that are challenging have been removed from the STEOM-CCSD evaluation set, making the statistical quantities possibly biased.  Nevertheless, the similarity between the 
STEOM-CCSD and CC2 performances observed in the present study was already underlined in earlier studies. \cite{Loo18a,Dut18,Loo20a,Loo20d} 

Moving now to methods with higher $\mathcal{O}(N^6)$ scaling, one notices that CCSD yields 
too large VTEs and a SDE comparable to the CC2 one. The overestimation trend of CCSD was again reported by various groups, though with magnitudes depending significantly on the actual test set, \cite{Sch08,Car10,Wat13,Kan14,Kan17,Dut18,Ver21} 
an aspect discussed below.  In contrast, ADC(3) delivers too small transition energies (MSE of $-0.10$ eV) and a rather large dispersion (SDE of $0.22$ eV). This relatively poor performance of ADC(3) is consistent with our findings for smaller 
compounds, \cite{Loo20b} but can be efficiently mitigated by averaging with the ADC(2) VTEs. Indeed, the error patterns of ADC(2) and ADC(3) are opposite, \cite{Loo18a} and the simple average between the results of these
two methods is effective: ADC(2.5), which has in practice the same computational cost as ADC(3), is the only  $\mathcal{O}(N^6)$ method delivering MSE, MAE, RMSE, and SDE all smaller than $0.10$ eV for the present set. 

To finish, we have assessed four CC methods including triples against the CCSDT-quality reference data for the singlet transitions.  The two perturbative approaches, CCSD(T)(a)* and CCSDR(3), deliver very similar statistical errors and can be essentially 
viewed as equivalent for practical purposes. Such similarity was reported by Matthews and Watson for very compact compounds, \cite{Mat16} but with significantly smaller deviations as compared to CCSDT.  Turning to 
the two CC approaches with approximated iterative triples, namely CCSDT-3 and CC3, the latter seems to have the edge for the present set, with average errors below the chemical accuracy threshold ($0.043$ eV). We note that a similar ranking was
reported in 2017 for small molecules. \cite{Kan17}
	
The central section of Table \ref{Table-7} reports MAEs obtained for various subsets. We underline that the CT and $\ps$ subsets are both limited in size (13 and 6 ESs) and nature (rather weak CT and transitions found in TTF only, respectively). Hence, the 
present trends should be analyzed cautiously.  Most tested approaches deliver rather similar MAEs for the singlet and triplet ESs. However, CCSD is clearly more accurate for the triplets, in line with their larger $\Td$ values than the singlets, whereas both
STEOM-CCSD and CIS(D) apparently show the opposite behavior with better performances for the singlets.  Surprisingly, most methods provide more accurate Rydberg VTEs than valence VTEs; the differences are especially large for
CIS(D), CCSD, ADC(3), ADC(2.5) and all CC models including triples with significantly smaller MAEs for the Rydberg transitions. Although such trend are not unprecedented, \cite{Ver21} the differences between valence and Rydberg
MAEs are quite large here.  Interestingly, for much small systems, Kannar, Tajti, and Szalay reported that CC2 was much less accurate for the Rydberg transitions, \cite{Kan17} an effect that we do not observe here. We believe that this might be related to the 
low-lying nature of the transitions considered here whereas the Rydberg transitions of  Ref.~\citenum{Kan17} correspond to high-energy ESs with a very diffuse character. For the present set of ESs, one also notices significant differences of MAEs for the 
$\np$ and $\pp$ transitions, the latter being typically more accurately treated, except with STEOM-CCSD.

Let us now turn towards the impact of the molecular size, so as to investigate if larger compounds are more or less challenging than smaller ones for the different methods. In the three rightmost columns of Table \ref{Table-7}, we compare the MAEs determined 
for the present set to those reported in the QUEST database for molecules containing 1--3 and 4--6 non-hydrogen atoms. \cite{Ver21}  One can broadly divide the various methods in three groups. In the first group, one notices a significant deterioration 
of the performance as the system size increases, hinting that the method cannot be reasonably recommended for large compounds. In this category, one finds: (i) EOM-MP2, with a MAE quadrupling between the smallest and 
largest molecules, likely explaining why this method gave reasonable performances in previous works; \cite{Taj16,Kan17} (ii) CCSD with an error steadily increasing from $0.07$ to $0.18$ eV as the size of the benchmarked compounds increases; 
and (iii) CCSD(T)(a)* and CCSDR(3) with MAEs tripling between the smallest and largest molecules, though remaining under the $0.10$ eV threshold. While the perturbative triples clearly help in correcting the CCSD VTEs, they are insufficient in order to 
fully compensate the CCSD overestimations for larger molecules.  In the second category of approaches, Table \ref{Table-7} indicates significant improvements of the accuracy as system size increases, i.e., these methods are likely recommendable 
for ``real-life'' applications. Both ADC(2) and CC2 belong to this category, with a MAE divided by a factor of two when considering the larger rather than smaller molecules. This trend likely explains the contrasted CC2 results
obtained previously when experimental references on large compounds were considered, \cite{Goe10a,Win13,Jac15b} or when high-level TBEs on tiny molecules were used \cite{Sch08,Kan17,Loo18a} as references.  In the last category, one 
finds approaches that behave relatively similarly for all system size, and this is essentially all the remaining methods. 

\section{Concluding remarks}

The vertical transition energies to 91 ESs of bicyclic molecules containing between 8 and 10 non-hydrogen atoms have been evaluated.  The selected molecules were chosen to be representative of $\pi$-conjugated chromogens actually present
in dyes and fluorophores used daily in measurements, e.g., azulene, benzothiadiazole, diketopyrrollopyrrole, quinoxaline, and tetrathiafulvalene.  To define our TBEs/{\AVTZ} reference values, we typically relied on CC3 for the triplet excited states 
as they are characterized by a strongly dominant single-excitation character. For the singlet ESs, we employed corrected CC3 VTEs thanks to double-$\zeta$ CCSDT values. Although the present set does not include transitions with a (dominant) 
double excitation character, it nevertheless includes a reasonable mix of singlet (58), triplet (33), valence (60), Rydberg (17), and CT (13) ESs. Such variety of states is rather typical for molecules of this size, though systems with stronger CT 
character are missing. \cite{Koz20,Loo21a}  

Thanks to these TBEs, we have benchmarked 16 wave function methods commonly employed for ES calculations on this type of systems. We stress that the present set considerably extends the  number of ESs of ``large'' molecules  (more than
6 non-hydrogen atoms) included in the QUEST database. \cite{Ver21} This allows for a fair evaluation of the performances of various methods as the size of the molecules increases.  The outcomes of the present study clearly show that the quality 
of the VTEs provided by both ADC(2) and CC2 improves with system size, whereas CCSD follows the opposite pathway. In the group of computationally ``cheap'' models, it seems reasonable to recommend ADC(2), CC2, and STEOM-CCSD. 
Besides, ADC(2.5) appears as a good compromise to further improve the accuracy: it outperforms the three previous methods almost systematically for a cost equivalent to ADC(3).  For the present set, ADC(2.5) delivers statistical deviations 
comparable to those obtained with the more resource-intensive CCSD(T)(a)*, CCSDR(3), and CCSDT-3 approaches, whereas in the case of smaller compounds, the reverse trend was clearly found. Finally, it is probably worth stressing that the 
present work hints that CC3 delivers excellent accuracies for  all treated subsets (ES nature, size of the compounds, etc).  While CC3's $\mathcal{O}(N^7)$ scaling prevents applications on large systems with large basis sets, CC3/double-$\zeta$ 
calculations are achievable on chromophores, fluorophores, and photochromes of actual practical interest, e.g., substituted naphthoquinones, coumarins, and azobenzenes. 

The present results could serve as a reliable guide to select a cheaper level of theory for specific classes of organic compounds. Such a purely theoretical strategy to select a low-scaling theoretical model (e.g., an adequate exchange-correlation 
functional to perform TD-DFT calculations), on the basis of a straightforwardly accessible property (the vertical transition energy in the present case) can likely be viewed as complementary to the protocols developed by Barone and coworkers 
in which they model complex spectroscopic properties (such as vibronic spectra) so as to allow direct comparisons with experiment. \cite{Egi13,Bai13,Bai18}

\section*{Acknowledgements}
PFL thanks the European Research Council (ERC) under the European Union's Horizon 2020 research and innovation programme (grant agreement no.~863481) for financial support. 
DJ is indebted to the CCIPL computational center installed in Nantes for (the always very) generous allocation of computational time. 
\section*{Supporting Information Available}
Raw CCSD/{\AVTZ} data. Cartesian coordinates. Theoretical best estimates. VTEs for all benchmarked methods.

\bibliography{biblio-new}

\end{document}